\documentclass[preprint,12pt] {elsarticle}

\usepackage{latexsym,amssymb,amsmath}

\begin{document}

\begin{frontmatter}


\title{Punctuated equilibrium dynamics in human communications}


\author[a]{Dan Peng}
\author[b]{Xiao-Pu Han}
\author[a]{Zong-Wen Wei}
\author[a,c,d]{Bing-Hong Wang}
\address[a]{Department of Modern Physics, University of Science and Technology of China, Hefei 230026, China}
\address[b]{Alibaba Research Center for Complexity Sciences, Hangzhou Normal University, Hangzhou 311121, China}
\address[c]{College of Physics and Electronic Information Engineering, Wenzhou University, Wenzhou 325035, China}
\address[d]{The Research Center for Complex System Science, University of Shanghai for Science and Technology, Shanghai, 200093 China}

\begin{abstract}
A minimal model based on individual interactions is proposed to study the non-Poisson statistical properties of human behavior: individuals in the system interact with their neighbors, the probability of an individual acting correlates to its activity, and all individuals involved in action will change their activities randomly. The model creates rich non-Poisson spatial-temporal properties in the activities of individuals, in agreement with the patterns of human communication behaviors.
Our findings provide insight into various human activities, embracing a range of realistic social interacting systems, particularly, intriguing bimodal phenomenons. This model bridges priority queues and punctuated equilibrium, and our modeling and analysis is likely to shed light on non-Poisson phenomena in many complex systems.
\end{abstract}


\end{frontmatter}



\section{Introduction}
\label{}
In the recent decade, along with the fast development of online social services, the understanding and predicting of human behaviors has attracted much attention of researchers. One of the remarkable features of statistical patterns of human behavior is the wide-spread non-Poisson properties \cite{Bar2005, Oliveira2005, Eckmann2004, Dez2006, Gon2008, Slijper2007, Radicchi2009, Chen2010, Takaguchi2011, JiangZQ2013}, which usually shows heavy-tail distribution on temporal statistics or spatial patterns and sharply differ from the Poissonian picture in traditional understanding \cite{Haight, Reynolds}.
Several mechanisms based on separated individuals have been proposed to explain the origin of bursts and heavy tails, including priority-queuing processes \cite{Bar2005, Vaz2006, Vaz2005, Gon2008}, Poisson processes modulated by circadian and weekly cycles \cite{Malmgen2008, Malmgren2009, Jo2012}, adaptive interests \cite{Gon2008, Han2008}, preferential linking \cite{Gon2008}, and memory effects \cite{Karsai2012}.

However, one major concern of behavior types in human dynamics is the interaction and communication behavior between individuals. Such human actions have a strong impact on resource allocation, circulation of information, even evolution of social network structure, therefore it have been the focus of research. In real life, everyone is influenced by the surrounding social environment, for instance, the interval between sending two consecutive E-mails is influenced by the actions of this individual and the other communicating partners.
The researchers achieved a breakthrough first in the simplest model of two interacting bodies \cite{Oliveira2009}. And a minimal model of interacting priority queues to discuss bimodal phenomenon observed in Short Message correspondence is proposed \cite{Wu2010}: inter-event time distribution was neither completely Poisson nor power law but a bimodal combination of them. Notably, this bimodal phenomenon was also observed in inter-event time distribution of the calling activity of mobile phone users \cite{Candia2008}, two consecutive transactions made by a stock broker \cite{Vaz2006} and successive transactions of experimental futures exchange \cite{WangSC2008} \emph{etc}. Indeed, some purported power-law distributions in complex systems may not be power laws at all. In fact, strict power-law distribution is rarely observed in empirical studies albeit bursts and heavy tails are widespread. In addition to the widespread bimodal distribution, human activity patterns may be power-law distribution followed by distinct cutoff \cite{Scherrer2008, JiangZQ2013}, multimodal distribution of power-law with different scaling exponent \cite{WangP2011, Grabowski2009, ChunH2008}, or in some instances it is more consistent with Mandelbrot distribution \cite{RenXZ2012}, and so on. Human activity patterns exhibit such a wealth of statistical properties. How to quantitatively understand human dynamics, dose there exist an universal fundamental governing human dynamics and individual? The understanding about these questions need to be studied deeply.

In this paper, a simple model incorporating solely individual interaction based on network is proposed to explain and develop human dynamics, especially the origin of bursts and heavy tails. Our model links punctuated equilibrium models and queueing models, and reproduces varieties of distributions on spatial-temporal patterns observed in empirical researches, \emph{e.g.}, exponential distribution, power-law distribution, and bimodal properties.

\section{The model}

In real life,
a certain type of activity of an individual is influenced by the actions of this agent and the other communicating partners. For example, one person may reply in no time after receiving a message, or even send one more to someone else; without receiving any message, however, this person may not send out any message in a long time. That is, most of the time, people's social behavior is activated by others. Whereas, it is conceivable that the interaction make individual passive, for instance, an ongoing topic between interacting individuals terminates, or one party is reluctant to keep up interacting and so on.
Hence we consider that social interactions make an individual more active, or more inactive.
Yet an activated individual might keep reticent or contact to more than one neighbor. On the other hand, an individual might act at will without environmental stimulation. Incorporating all these considerations, the schemes of our model are as follows:

$N$ nodes (individuals) are arranged on a network. We define a series of massage sendings that are activated by a common source node as a ``burst".
Each node may be in two different status: affected by a burst (A), or un-affected status (U).
Initially ($t = 0$), each node (for node $i$, say) is in U status, and its activity value is set as a random number $a_i$, equally distributed between 0 and 1. At each processing step $t$, the evolution of the system obeys the following rules:

i) with probability $p$, the node with the highest activity is chosen to send messages to its $n$ neighbors; or with probability $1-p$, an arbitrary node is chosen to be the sender, denoted as $S_{t}$. Here $n$ is the number of messages sent in one processing step and is not larger than the degree of the sender;

ii) If the sender is in status U, namely, the current message sending is out of the previous burst, so all the other nodes turn to U status.

iii) If the sender has received message in previous time steps, with probability $q$, the sender sends a message back to the node which sent the last massage to the sender. Or with probability $1-q$, it sends a message to a randomly selected neighbor of sender $S_{t}$, and sending other $n-1$ messages to neighbors randomly.

iv) The sender and all the nodes received message at the current time step turn to A status to start a new burst, and update their activity values to be new random numbers between 0 and 1.

v) Go into the next step $t+1$ and repeat the above procedures.



This model has three free parameters for a given network: $n$, $p$ and $q$.
The interaction come about when messages are sent ($n \geq 1$) by the most active individual ($p = 1$) and all individuals involved in the action mutate their activities. When $p < 1$, it incorporates Poisson initiation of messages. If the network is regular (for example, a Ring) and $n$ equals the degree of nodes, this simple model is same as Bak-Sneppen model. If $n = 0$, the direct interaction between individuals is ignored completely, this case is equivalent to Barab\'{a}si queueing model. Thus we give an association between priority-queue models and criticality phenomena.

\section{Simulation Results}

\subsection{Simulations on Ring}

\begin{figure}
   \includegraphics[width=12cm]{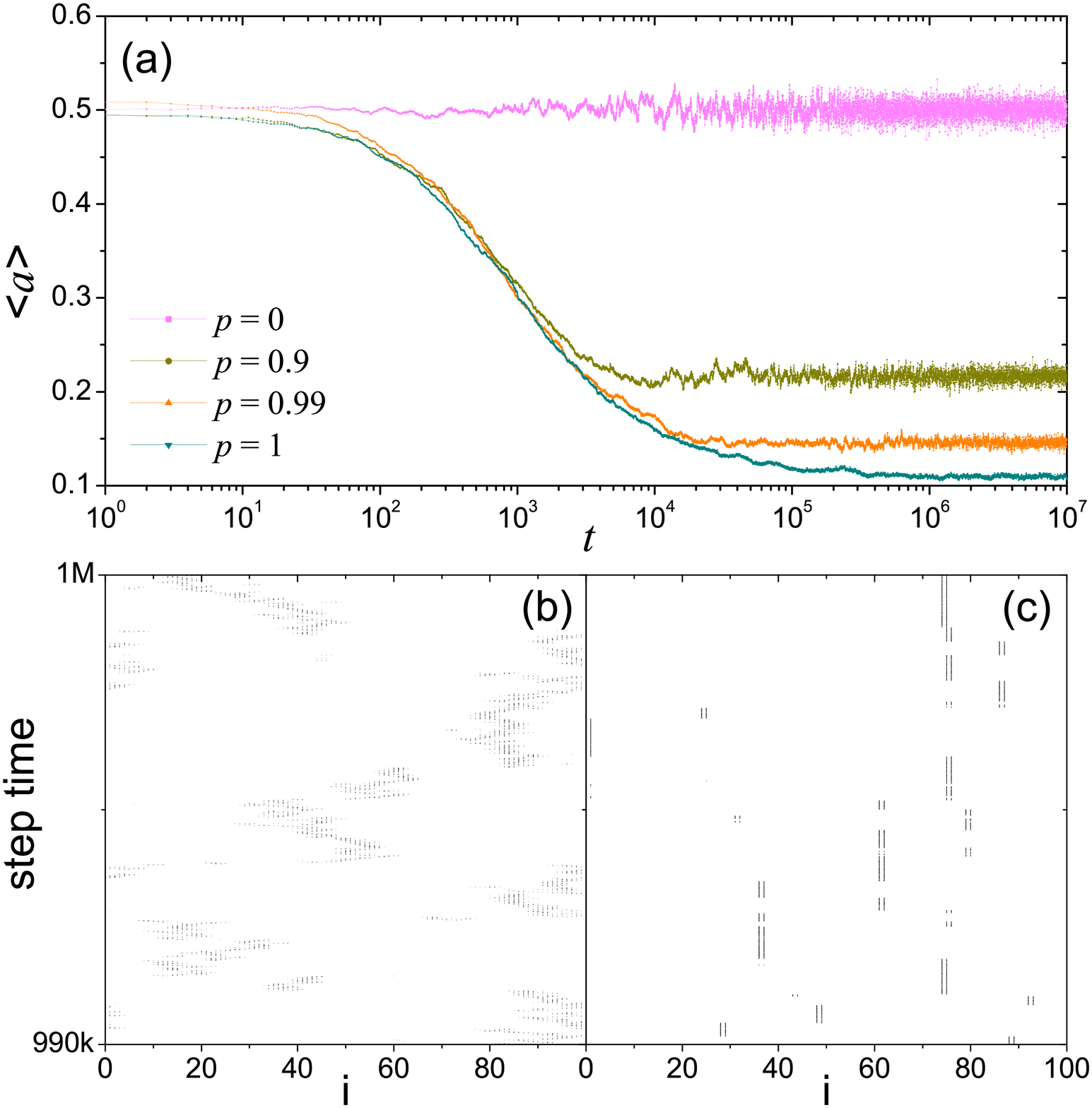}\\
  \caption{(Color online)
  (a) The evolution of the global average activity $\langle a \rangle$ for different $p$, and the model runs on a Ring with parameters $N=10^3$, $n=1$, $q=0$.
  (b) and (c) Spatial-temporal patterns of the model running on a Ring in the stable period under the parameters setting $N=100$, $p=0.999$, $n=1$ and different $q$.
  (b) $q=0$, the system emerges long-range correlation similar to the result of BS model. (c) $q=1$, message usually propagates between a pair of nodes. }
\end{figure}

In the first instance, we concentrate on the dynamics on a Ring that every individual has two neighbors.
It has several modes for the sending of messages. The first one is that the sender send no message to its neighbor, namely, $n=0$, and the interaction between individuals is eliminated. This case is equivalent to Barab\'{a}si queueing model, in which the node with highest activity is updated with probability $p$.
The second mode is that the sender send messages to all neighbors, namely, $n = k$. For Ring ($n=2$), when $p = 1$, the model is same as Bak-Sneppen model and exhibits long-range correlation and punctuated equilibrium (see Fig. 1(a) and (b)).
Decreasing the value of $p$, along with the growth of randomness, the long-term correlation weaken and the stable value of global average activity $\langle a \rangle$ increases (see Fig. 1(a)).
When $p=0$, sender is completely randomly-chosen, and the system is in a disordered state with $\langle a \rangle = 0.5$ (see Fig. 1(c)).
The third mode corresponds to the case between the former two modes, namely, $0<n<k$ (for Ring, the only case is $n=1$), which is impacted by the selection of receivers (decided by $q$). In the extreme case $q=1$, massage sendings usually are restricted to one pair of individuals, as shown in Fig. 1(c).  

In the following we meticulously investigate statistical patterns on a Ring for different $p$ and $q$. For simplicity of discussion, we fix $N = 10^3$ and $n = 1$. Here all distributions in the following discussions are counted in stable period.

Firstly we investigate inter-event time distribution $P(\tau)$, here the inter-event time $\tau$ is defined as the time interval between two consecutive massage sendings of a node. For simplicity of discussion about $p$, we fix $q$ value, in particular $q = 0$. As $p = 0$, at each step, sender is selected randomly, $P(\tau) = \frac{1}{N}(1-\frac{1}{N})^{\tau-1}$.
In the case $p=1$ (BS model), the system evolves into a typical status of self-organization criticality and $P(\tau)$ obeys power-law distribution with exponent $-1.56$ (Fig. 2(a)).
With the decrease of $p$, the long-range correlations in each avalanche range is partially broken by the randomness of sender selection, an exponential tail on $P(\tau)$ emerges and stretch $P(\tau)$ into bimodal form (Fig. 2(a)),
corresponding to long intervals of uncorrelated message sendings.
 And when $p = 0$, the system is completely Poissonian-like (Fig. 2(a)).

By contrast, the impact of $q$ on $P(\tau)$ is not so strong. As shown in Fig. 2(b), when $p$ is very close to 1, as  $q$ increases from $0$ to $1$, a small Poissonian-like region appears in the head of $P(\tau)$, and the tail keeps power-law-like form.


The waiting time $\tau_w$ is the time interval between the message sending and the last receiving of a node. The results of $P(\tau_w)$ are very similar to $P(\tau)$, as shown in Fig. 2 (c) and (d).

Meanwhile, the spatial distribution $P(d)$ of the distance $d$ between the senders of two consecutive time steps, is deeply impacted by both $p$ and $q$.
Fig. 2(e) shows the correlated range (the power-law region of $P(d)$) enlarges from zero to the global range as $p$ increases from $0$ to $1$.
And localized interactions ($q>0$) weaken the long-range correlation (see Fig.2(f)).

\begin{figure}
  \includegraphics[width=12cm]{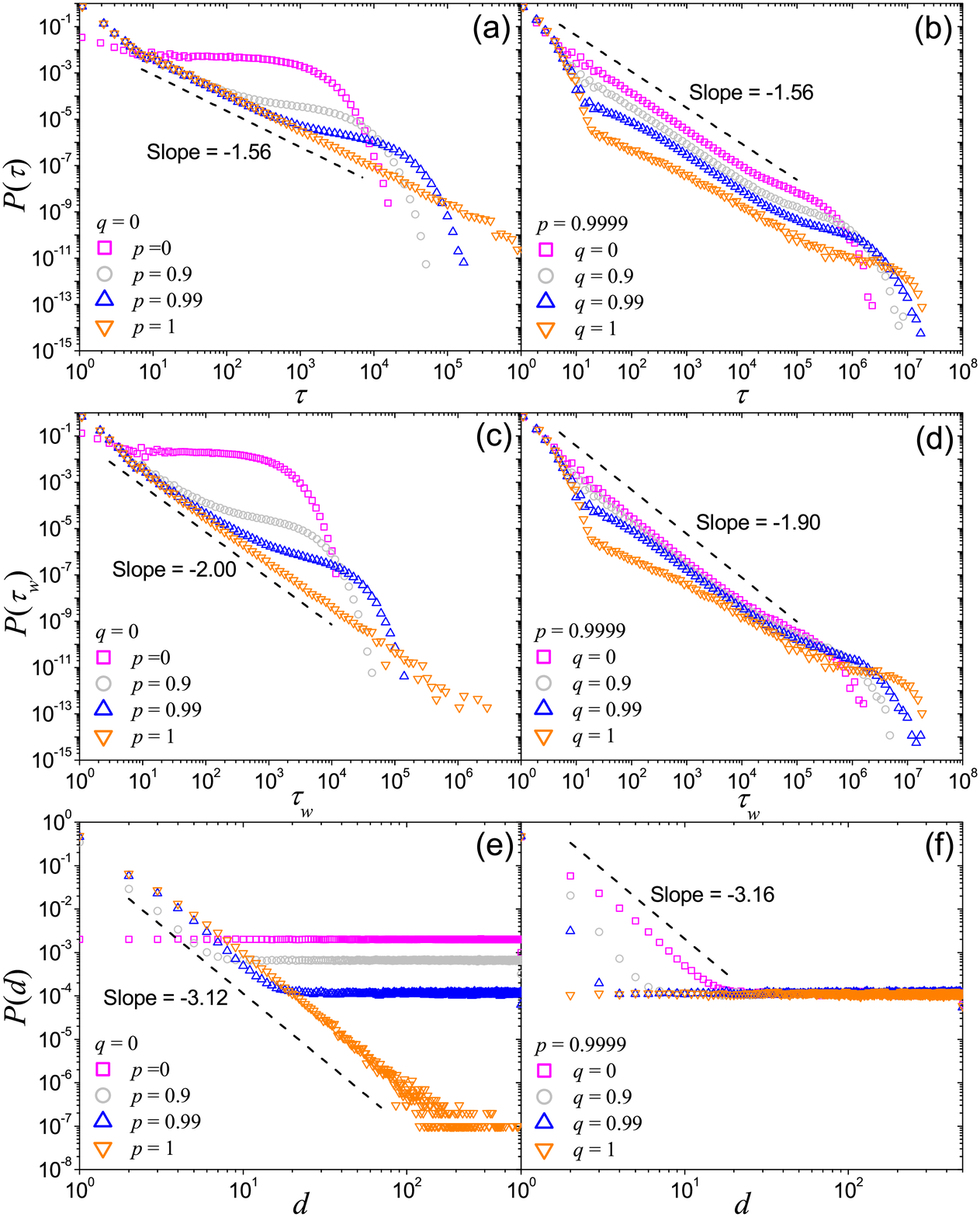}\\
 \caption{(Color online) The distribution $P(\tau)$ for (a) different $p$ with $q = 0$ and (b) different $ q $ with $ p=0.9999 $.
 The distribution of waiting time $P(\tau_w)$ for (c) different $p$ with $q = 0$ and (d) different $q$ with $p=0.9999$.
 The distribution $ P(d) $ for (e) different $p$ with $q = 0$ and (f) different $q$ with $p=0.9999$. Other parameters are fixed as $N=10^3$ and $n=1$. The black dash lines denote the power laws with the corresponding exponents. }
\end{figure}


\begin{figure}
  \includegraphics[width=12cm]{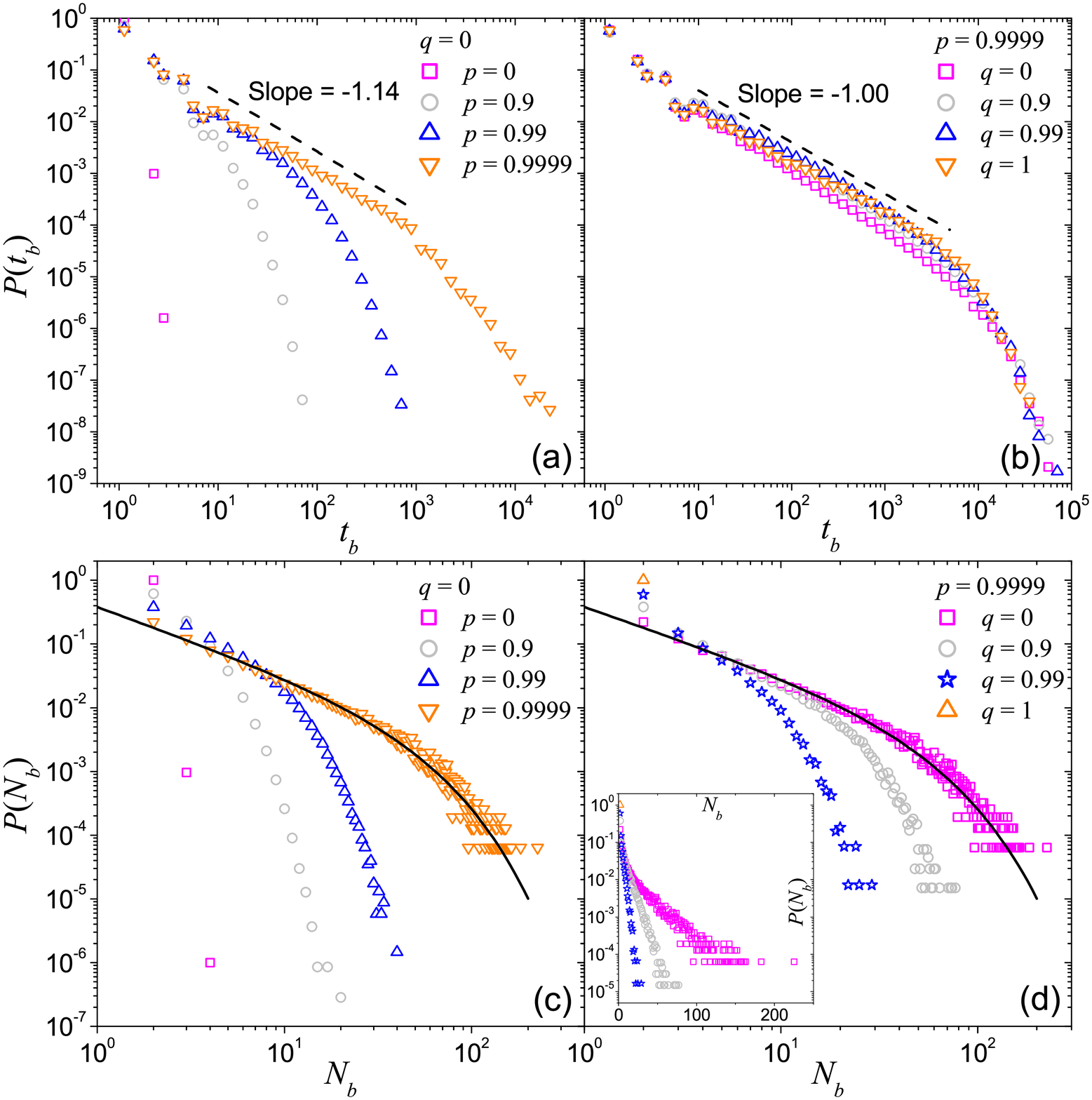}\\
  \caption{(Color online) The distribution $P(t_{b})$  for (a) different $p$ with $q=0$ and (b) different $q$ with $p=0.9999 $;
  the distribution $P(N_{b})$ for (c) different $p$ with $q=0$ and (d) different $ q $ with $ p=0.9999 $. Other parameters are fixed as $ N=10^3 $ and $ n=1 $. The dash lines show the fitting curves. And in panel (c) and (d), the fitting function is $P(N_b)=45N_b^{-1.05}\exp{[-(0.025x+4.75)]}$.
  The solid lines denote the power laws with the corresponding exponents.}
\end{figure}

Moreover, the spatial-temporal patterns of bursts also show similar scaling properties.
The burst duration time $t_b$ that is defined as the total number of time steps in a burst period,
is expected to be 1 as $p = 0$, and follows power-law-like distribution while $p$ value very closes to 1.
This power-law-like distribution of burst sizes was observed in the distribution of Orkut session \cite{Benevenuto2012} and the spread of news and opinions \cite{Leskovec2009}. It is interesting that $P(t_{b})$ is strongly sensitive to the reduction of $p$, seemingly follows exponential distribution (see Fig. 3(a)), implying that scaling temporal pattern results from priority-activity mechanism of acting in system.
In contrast, the setting of $q$ almost can not impact on $P(t_b)$ (Fig. 3(b)).
Ref. \cite{JiangZQ2013} reported various distribution patterns of burst sizes illustrated in Figs. 3(a) and (b), such as exponential, power law and even a distribution similar to $P(t_{b})$ at $p=0.99$.
Nevertheless, the spatial property of bursts is slightly different.
As shown in Fig. 3(c) and (d), when $p\rightarrow1$ and $q\rightarrow0$, the distribution $P(N_b)$ obeys power law with exponential cutoff, here $N_b$ is the affected range of burst and is defined as the number of nodes involved in the burst. In contrast with $P(t_b)$, besides $p$, $P(N_b)$ also shows sensibility on $q$: large reply probability $q$ will break the scaling property on $P(N_b)$ (see Fig. 3(d)).
When $p=1$, there usually exists only one burst.

\subsection{Topological effects}

Topology of media usually deeply affect the dynamics of system. In this section, we run the model on different modeling networks to investigate the topological effect on the process.

The above discussions are based on the dynamics on regular networks. With the increase in the probability $r$ of randomly rewriting edges, the average distance of the network rapidly reduces, and regular network changes to be small-world network. So, we firstly run the model on three types of media: Nearest-neighboring regular network, Watts-Strogatz (WS) small-world network with different rewriting probability $r$, and Erd\:{o}s-R\'{e}nyi (ER) random network. All the networks have the same size ($N=10^4$) and the same average degree ($\langle k\rangle=4$), and the model runs on the typical parameter setting leading to scaling property: $n=1$, $p=1$, $q=0$.

\begin{figure}
  \includegraphics[width=13.8cm]{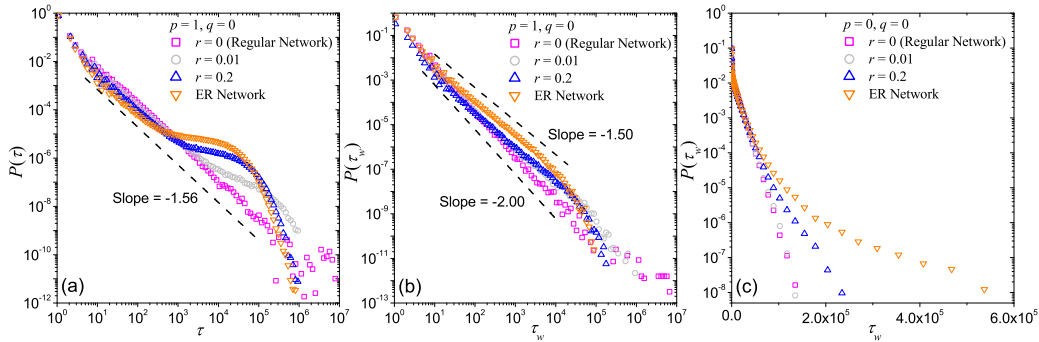}\\
  \caption{(Color online) Effect of topological randomness. The model runs on WS networks for different rewriting probability $r$ and ER random network.  (a) the inter-event time distribution $P(\tau)$, parameter settings $p=1$, $q = 0$; (b) the waiting time distribution $p(\tau_w)$, parameter settings $p=1$, $q = 0$; (c) the waiting time distribution $p(\tau_w)$, parameter settings $p=0$, $q = 0$. Other parameters are set on $N = 10^4$, $n=1$. Dashed lines show the fitting slopes of curves. }
\end{figure}

Fig. 4(a) shows the differences on the inter-event time distribution $P(\tau)$ for these networks. Topological randomness mainly affects $P(\tau)$ in its tail:
 it drives the second peak rising on the tail of $P(\tau)$ but almost has no effect on the power-law head, implying another possible origin of the bimodal property would be the small diameter of network, because the long-range correlation in the critically is limited by the diameter of the network.
In contrast, the impact of the randomness on the waiting time distribution is weak. $P(\tau_w)$ on small-world networks and ER networks generally keeps power-law-like form and only has a slightly change on the power law exponents (see Fig. 4(b)).

\begin{figure}
  \includegraphics[width=13.5cm]{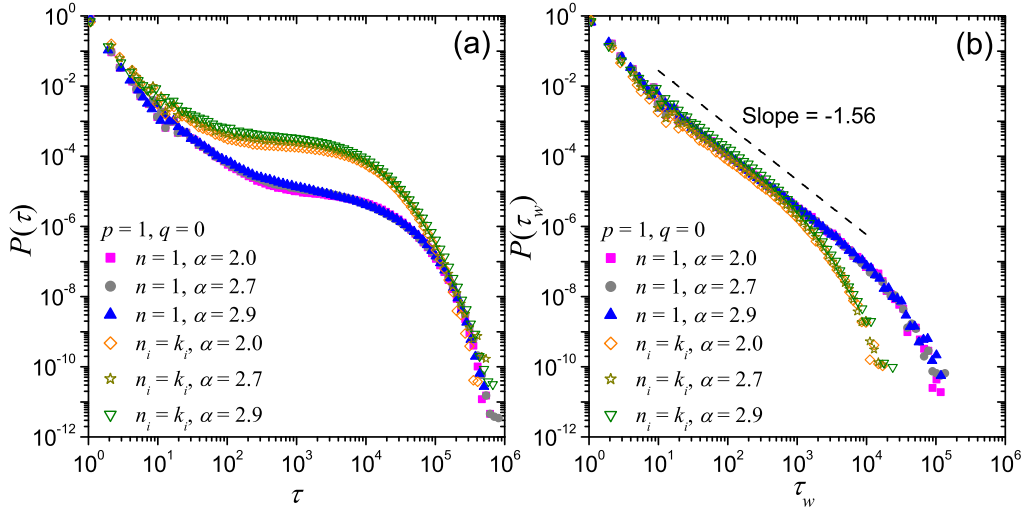}\\
  \caption{(Color online) Effect of heterogeneity of network degree. The model runs on random scale-free networks with different power law exponent $\alpha$ of degree distribution. (a) the inter-event time distribution $P(\tau)$ for different $\alpha$ and different $n$, parameter settings;
  (b) the corresponding waiting time distribution $P(\tau_w)$. Other parameters are set as $p=1$, $q = 0$, $N = 10^4$. Dashed lines show the fitting slopes of curves.
  }
\end{figure}

In addition, a noticeable case is that, when $p=0$ and $q=0$, topological randomness has contributions on the emergence of heterogeneous waiting time distribution. As shown in Fig. 4(c), the tail of $P(\tau_w)$ on ER random network decays much slower than the one on other networks and obviously deviates the exponential form. In this case, the punctuated equilibrium does not exist, and thus this effect would be purely related to topological impacts.

Furthermore, the simulation results on different random scale-free networks are compared to investigate the impact of heterogeneous topology. Due to the heterogeneity on degree distribution, we consider two cases on the setting of parameter $n$. The first one is $n=1$. And another one is $n_i = k_i$, namely, the sender broadcasts messages to all neighbors. Simulation results on the random scale-free networks with different power law exponent $\alpha$ of degree distribution are shown in Fig. 5.
Generally, the inter-event time distributions $P(\tau)$ on random scale-free networks are in bimodal type, and the second peak mainly depends on the value of $n$: broadcasting messages to all neighbors will leads to more higher peak. However, the power law exponent $\alpha$ of degree distribution almost has no effect on $P(\tau)$ (Fig. 5(a)), indicating that exponent of power-law degree distribution dose not play the dominant role in the dynamics of the system, which is sharply different to many other network-based dynamics. Similar phenomena is also observed on the waiting time distribution $P(\tau_w)$ (Fig. 5(b)).

\subsection{Simulations on real-world social networks}

\begin{figure}
  \includegraphics[width=13.5cm]{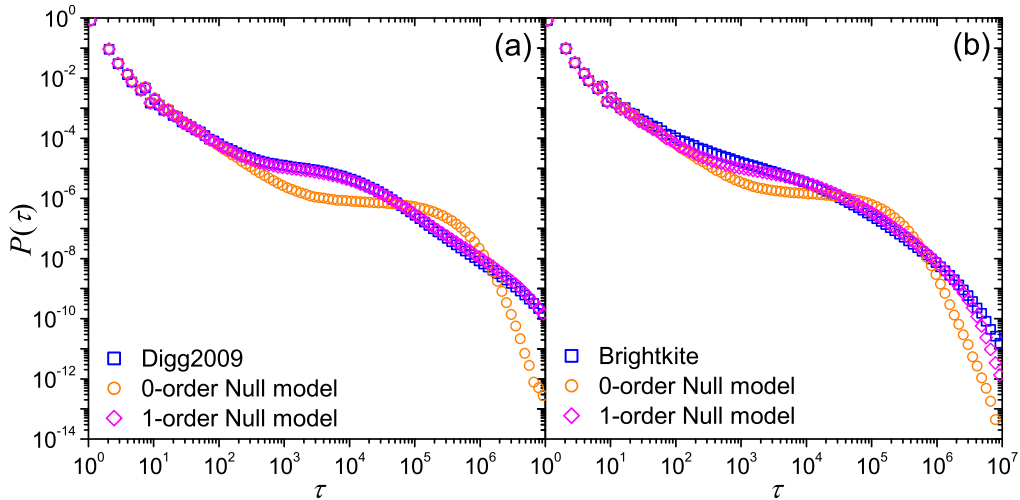}\\
  \caption{(Color online) Simulation results on two real-world social networks: (a) Digg2009 and (b) Brightkite. The three curves in each panel shows $P(\tau)$ of the original friendship network, the corresponding zero-order Null model and the one-order Null model, respectively. Simulations runs on parameter settings $p=1$, $q=0$, $n=1$.
  }
\end{figure}

We also run the model on real-world media. The datasets of real-world social networks are the friendship network of Digg users and  Brightkite users, respectively.
The detailed information of the real-world social networks can be found in {\it Appendix}. Both of the two networks are scale-free like.

Under the typical parameter setting $n=1$, $p=1$, $q=0$, as shown in Fig. 6, both of the two inter-event time distributions $P(\tau)$ on the real-world media are bimodal like. Nonetheless, the tail of $P(\tau)$ decays slowly and is close to power functional form, in contrast to the scenario on random scale-free networks shown in Fig. 5(a). Analogous bimodal phenomenon was observed in the inter-event time distribution of on-line book-marking \cite{WangP2011}.
In order to dig out the potential source of this difference, we compare the results on the original friendship networks with the cases on zero-order Null model and one-order Null model of the networks.

Null model is an usual method to investigate impact of the network structures on study.
A zero-order Null model is almost completely different with the original network except for the same numbers of nodes and edges. We construct Zero-order Null model by the following method£ºat each step,
a randomly selected edge is removed and then we randomly select two nodes,
add an edge if they are unconnected, repeat this procedure enough times. And one-order Null model is defined as that, the degree of each node remains unchanged, but links are randomly reassigned.
The method of constructing one-order Null model is: at each step, we randomly select two edges (edges $E_{ij}$ which connect nodes $i$ and $j$, and $E_{lm}$, say) which are do not possess same node, delete this two edges, add two new edges $E_{il}$ and $E_{jm}$ if neither $E_{il}$ nor $E_{jm}$ does not exist, and repeat this procedure until all edges are treated.

As shown in Fig. 6, on both of two datesets, $P(\tau)$ of one-order Null model is very close to the results on the original networks, indicating that the differences on degree distribution drives the difference on the modeling results between the the scenario on random scale-free networks and the one on real-world media.

\section{Discussion}

Taking account into the activation effect in social interactions, our model bridges queueing models and criticality phenomena, obtains rich statistical properties consistent with empirical studies.
The results of the model mainly includes the following points. One is rich non-Poisson properties in individual's activities.
, which corresponds to human activities in communications.
Noticed that, since in the model only one event occurs  in each time step, the time interval effectively is the interval of event. Therefore all the results in temporal patterns are in the mean of the ``relative clock" proposed in Ref. \cite{Zhou2012} that eliminates the effect of season fluctuation. Another point is the scaling patterns of bursts, which highly relates to many information spreading processes, such as the spreading range of meme and duration time of rumor.

In view of similar bimodal property is widely observed in many natural systems,
our model is likely to provide a new perspective on the bimodal property in earthquakes \cite{Parsons2002,Touati2009}, as well as other analogous natural systems, such as tsunami \cite{Geist2008}, rainfall \cite{Heneker2001}, and forest fire \cite{Benavent2007}.

In summary, the results of our model imply that, the punctuated equilibrium dynamics in human social interactions would be an important mechanism that drives the emergence of non-Poisson properties in human social activity patterns and many social dynamics. Although we arrange the scheme based on the considerations of human communications, with suitable modification, our model could be readily applicable to other interacting social systems such as trading and other economic systems.
This model would shed light on the studies on human communications and social dynamics, such as information spreading, rumor propagation and disease outbreak \cite{Small2007}.

\section{Acknowledgments}
\label{}
This work was funded by the National Important Research Project (Grant No. 91024026), the National Natural Science Foundation of China (No. 11205040, 11105024, 10975126, 11275186, 61403421), the Major Important Project Fund for Anhui University Nature Science Research (Grant No. KJ2011ZD07) and the Specialized Research Fund for the Doctoral Program of Higher Education of China (Grant No. 20093402110032).

\appendix

\section{Introduction of datasets and social networks}

\begin{figure}
  \includegraphics[width=13.5cm]{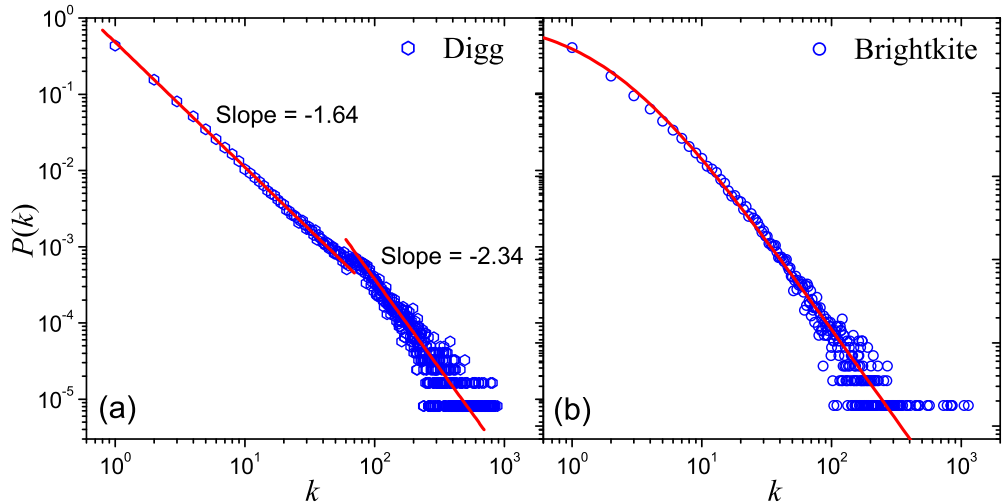}\\
  \caption{(Color online) The degree distributions of the two real-world social networks: Digg (a) and Brightkite (b). The fitting curve in panel (b) is the function $P(k)=3.8(x+2.0)^{-2.2}$.
  }
\end{figure}

One of the real-world social networks in this work is the Giant component which contains 702,995 friendship links of 154,513 users, extracted from Digg2009 dataset which contains data about stories promoted to Digg's front page over a period of a month in 2009. The dataset is available from the website http://www.isi.edu/$\scriptsize{\sim}$lerman/downloads/digg2009.html.
For each story, it collects the list of all Digg users who have voted for the story up to the time of data collection.
We also retrieved the voters' friendship links. We denote $k$ to be the number of friendship links of a user, and $P(k)$ to be the distribution of k. $p(k)$ is a power-law regime $P(k)=k^{-\alpha}$ with $\alpha_{1} =1.64$ for low $k$ and $ \alpha_{2} =2.34 $ for high $k$ (see Fig. 7 (a)).

The second real-world social network is the giant component which contains 212,950 friendship links of 56,712 users, extracted from Brightkite, which contains a total of 4,491,143 checkins of 58,228 users over the period of Apr. 2008 - Oct. 2010.The dataset can be downloaded from the website http://snap.stanford.edu/data/loc-brightkite.html. The friendship network was collected using their public API, and was constructed with undirected edges when there is a friendship in both ways. $P(k)$ follows ``shifted power law'' (see Fig. 7(b)).






\bibliographystyle{elsarticle-num}







\end{document}